\documentclass{xjkas}

\def\year{2014} % publication year
\def\month{August} % publication month
\def\volume{47} % journal volume
\def\beginpage{159} % first page of article
\def\endpage{161} % last page of article
\def\received{August 11, 2014} % date paper was received by JKAS
\def\accepted{August 22, 2014} % date of acceptance
\date{Received \received; accepted \accepted}

\newcommand\name[1]{{\small\sc #1}}
\def\PB{{P_{\rm B}}}
\def\Pj{{P_{\rm jet}}}
\def\MB{{\dot{M}_{\rm B}}}

\title{
Does the Jet Production Efficiency of Radio Galaxies\\ Control Their Optical AGN Types?\thanks{\sc Rapid Communication}
}

\author{Sascha Trippe}
\affil{Department of Physics and Astronomy, Seoul National University, 599 Gwanak-ro, Gwanak-gu, Seoul 151-742, Korea; \email{trippe@astro.snu.ac.kr}}
\def\corrauthor{
S. Trippe
}

\def\runningauthor{
Trippe
}

\def\runningtitle{
Jet Production Efficiency of Radio Galaxies vs. Optical AGN Type
}

\def\keywords{
galaxies: active; jets; nuclei; Seyfert --- accretion, accretion disks
}

\def\abstracttext{
The jet production efficiency of radio galaxies can be quantified by comparison of their kinetic jet powers $\Pj$ and Bondi accretion powers $\PB$. These two parameters are known to be related linearly, with the jet power resulting from the Bondi power by multiplication with an efficiency factor of order 1\%. Using a recently published \citep{nemmen2014} high-quality sample of 27 radio galaxies, I construct a $\PB-\Pj$ diagram that includes information on optical AGN types as far as available. This diagram indicates that the jet production efficiency is a function of AGN type: Seyfert 2 galaxies seem to be systematically (with a false alarm probability of $4.3\times10^{-4}$) less efficient, by about one order of magnitude, in powering jets than Seyfert 1 galaxies, LINERs, or the remaining radio galaxies. This suggests an evolutionary sequence from Sy\,2s to Sy\,1s and LINERs, controlled by an interplay of jets on the one hand and dust and gas in galactic nuclei on the other hand. When taking this effect into account, the $\PB-\Pj$ relation is probably much tighter intrinsically than currently assumed. 
}

\begin{document}
\jkashead %% set title, authors, abstract, etc.

%%%%%%%%%%%%%%%%%%%%%%%%%%%%%%%%%%%%%%%%%%%%%%%%%%%%%%%%%%%%%%%%%%%%%
%%% BEGIN MAIN TEXT HERE %%%%%%%%%%%%%%%%%%%%%%%%%%%%%%%%%%%%%%%%%%%%
%%%%%%%%%%%%%%%%%%%%%%%%%%%%%%%%%%%%%%%%%%%%%%%%%%%%%%%%%%%%%%%%%%%%%

\section{Introduction \label{sec:intro}}

Active galactic nuclei (AGN; see, e.g., \citealt{netzer2013} for a recent review) are supposedly powered by accretion of gas onto supermassive black holes located in the centers of probably all galaxies. A frequent, albeit as yet poorly understood, consequence of this process is the formation of collimated outflows, jets, which are emitters of synchrotron radiation especially prominent at radio wavelengths (see, e.g., \citealt{boettcher2012} for a recent review). One road to a better understanding of jet formation and the interplay of jets and accretion flows is provided by analysis of the energetics of the jets of nearby ($\lesssim$100\,Mpc) radio galaxies which have accretion rates much smaller ($\ll$1\%) than their Eddington limits. X-ray observations of those systems provide independent estimates of (i) the kinetic jet powers $\Pj$ and (ii) the Bondi accretion powers $\PB=\MB c^2$, with $\MB$ being the Bondi accretion rate and $c$ denoting the speed of light. Multiple studies \citep{allen2006,balmaverde2008,russell2013,king2013,nemmen2014,fujita2014} concluded that, within errors, $\Pj$ is proportional to $\PB$. Accordingly, one finds a jet production efficiency $\eta$ from $\Pj=\eta\,\PB$; observationally, $\eta\sim1$\% (ensemble average). However, the observed $\PB-\Pj$ relations show substantial ($\gtrsim$0.5 dex) intrinsic scatter, indicating an as yet overlooked systematic effect. This article addresses a possible cause.

\section{Data and Analysis \label{sec:data}}

I make use of the data set published recently by \citet{nemmen2014} [their Table~1] which has in turn been drawn from data by \citet{balmaverde2008} and \citet{russell2013} and which is marked by very careful selection. Jet and Bondi powers were derived from X-ray observations obtained by the Chandra space telescope. Kinetic jet powers were estimated from the ages and creation energies of jet-inflated X-ray cavities. Bondi accretion rates were estimated from temperature and density profiles of hot diffuse gas. (Please refer to \citealt{nemmen2014} for technical details.)

I collected optical AGN types from the NASA/IPAC Extragalactic Database (NED); the SIMBAD Astronomical Database; \citet{ho1997}; and \citet{veron2010}. Eventually, I found optical types for 17 of my 27 targets: two are of type Seyfert 1, six of type Seyfert 2, and nine of type Low-Ionization Nuclear Emission-Line Region galaxy (LINER -- cf. \citealt{ho2008}). The final dataset is summarized in Table~\ref{tab:data}.

From the data of Table~\ref{tab:data}, I constructed a logarithmic $\PB-\Pj$ diagram (Figure~\ref{fig:powers}) that includes the available information on optical AGN types. The sample spans about three orders of magnitude in both jet power and accretion power. I determined the best-fit linear relation between jet and accretion power via a least-squares fit, leading to a (ensemble averaged) jet production efficiency of $\eta\approx0.6$\% for my sample. The intrinsic scatter (i.e., the difference in squares of r.m.s. residual and bivariate r.m.s. measurement error) is about 0.5 dex.

%%% TABLE %%%%%%%%%%%%%%%%%%%%%%%%%%%%%%%%%%%%%%%%%%%%%%%%%%%%%%%%%%%%%%%%%%%%%
\begin{table}[t!]
\caption{The radio galaxy sample.\label{tab:data}}
\centering\vskip-4pt
\renewcommand\arraystretch{1.2}
\begin{tabular}{llll}
\toprule
Galaxy & Type & $\log\PB$ [W] & $\log\Pj$ [W] \\
\midrule
3C 066B    & Sy\,1   & $38.59_{-0.16}^{+0.12}$ & $37.25\pm0.40$ \\
3C 083.1   &         & $39.11_{-0.13}^{+0.10}$ & $36.69\pm0.40$ \\
3C 270     & L\,2    & $38.41_{-0.16}^{+0.12}$ & $36.77\pm0.40$ \\
3C 296     &	    & $39.36_{-0.20}^{+0.14}$ & $37.05\pm0.40$ \\
3C 449     &	    & $38.48_{-0.21}^{+0.14}$ & $36.66\pm0.40$ \\
3C 465     & Sy\,1   & $39.15_{-0.12}^{+0.10}$ & $37.58\pm0.40$ \\
UGC 6297   & Sy\,2   & $36.68_{-0.31}^{+0.18}$ & $34.83\pm0.40$ \\
UGC 7386   & L\,1    & $37.95_{-0.24}^{+0.16}$ & $36.18\pm0.40$ \\
UGC 7898   &         & $37.81_{-0.12}^{+0.09}$ & $35.53\pm0.40$ \\
UGC 8745   & Sy\,2   & $38.62_{-0.26}^{+0.16}$ & $35.78\pm0.40$ \\
NGC 1399   & Sy\,2   & $38.45_{-0.14}^{+0.11}$ & $35.34\pm0.40$ \\
NGC 3557   &         & $38.02_{-0.23}^{+0.15}$ & $35.87\pm0.40$ \\
IC 1459    & L       & $38.48_{-0.12}^{+0.09}$ & $36.89\pm0.40$ \\
IC 4296    & L       & $39.23_{-0.10}^{+0.08}$ & $36.94\pm0.40$ \\
NGC 6166   &         & $37.60_{-0.30}^{+0.18}$ & $35.95_{-0.11}^{+0.16}$ \\
NGC 4696   & L       & $36.71_{-0.05}^{+0.05}$ & $36.15_{-0.15}^{+0.18}$ \\
HCG 62     &         & $37.70_{-0.22}^{+0.15}$ & $35.78_{-0.18}^{+0.22}$ \\
M 84       & Sy\,2   & $38.11_{-0.03}^{+0.03}$ & $35.04_{-0.20}^{+0.26}$ \\
M 87       & L\,2    & $39.30_{-0.10}^{+0.08}$ & $35.90_{-0.20}^{+0.27}$ \\
M 89       & Sy\,2   & $37.70_{-0.70}^{+0.26}$ & $34.48_{-0.18}^{+0.22}$ \\
NGC 507    &         & $38.30_{-0.30}^{+0.18}$ & $36.28_{-0.20}^{+0.24}$ \\
NGC 1316   &         & $37.70_{-0.70}^{+0.26}$ & $34.95_{-0.26}^{+0.28}$ \\
NGC 4472   & Sy\,2   & $37.95_{-0.18}^{+0.12}$ & $34.85_{-0.24}^{+0.23}$ \\
NGC 4636   & L\,1    & $36.70_{-0.10}^{+0.08}$ & $34.43_{-0.18}^{+0.19}$ \\
NGC 5044   &         & $37.30_{-0.30}^{+0.18}$ & $35.11_{-0.16}^{+0.21}$ \\
NGC 5813   & L\,2    & $37.00_{-0.22}^{+0.15}$ & $34.85_{-0.24}^{+0.27}$ \\
NGC 5846   & L\,2    & $37.11_{-0.64}^{+0.25}$ & $35.20_{-0.20}^{+0.24}$ \\
\bottomrule
\end{tabular}
\tabnote{{\sc Notes:} ``Type'' refers to optical AGN type; Sy\,1/2: Seyfert 1/2 galaxies; L: LINERs; L\,1/2: type 1/2 LINERs.
\\ {\sc References:} Source types: NED; SIMBAD; \citet{ho1997,veron2010}. Powers: \citet{balmaverde2008,russell2013,nemmen2014}.}
\end{table}
%%%%%%%%%%%%%%%%%%%%%%%%%%%%%%%%%%%%%%%%%%%%%%%%%%%%%%%%%%%%%%%%%%%%%%%%%%%%%%%

\section{Results and Discussion \label{sec:discuss}}

%%% FIGURE %%%%%%%%%%%%%%%%%%%%%%%%%%%%%%%%%%%%%%%%%%%%%%%%%%%%%%%%%%%%%%%%%%%%
\begin{figure}
\includegraphics[angle=-90,width=82mm]{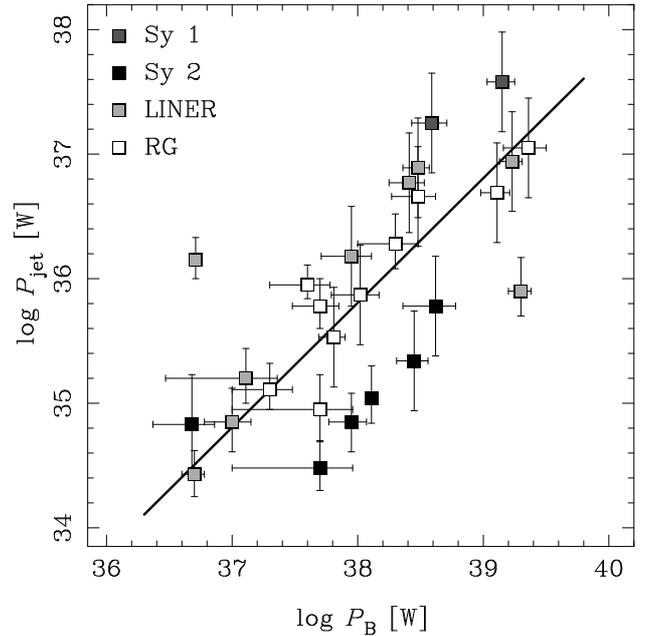}
\caption{Kinetic jet power $\Pj$ as function of Bondi accretion power $\PB$ using the data in Table~\ref{tab:data}. Optical AGN types are Seyfert 1 (Sy\,1), Seyfert 2 (Sy\,2), and LINER; RG denotes radio galaxies of unknown type. Error bars indicate $1\sigma$ uncertainties. The diagonal continuous line corresponds to a linear relation with a jet production efficiency of $\eta\approx0.6$\%. \label{fig:powers}}\vskip-5pt
\end{figure}
%%%%%%%%%%%%%%%%%%%%%%%%%%%%%%%%%%%%%%%%%%%%%%%%%%%%%%%%%%%%%%%%%%%%%%%%%%%%%%%

Inspection of Figure~\ref{fig:powers} reveals that the sample galaxies follow a linear relation between $\PB$ and $\Pj$ globally, albeit with substantial systematic scatter. As already noted by \citet{allen2006}, the presence of this relation, as well as the value found for $\eta$, indicate (i) that the accretion flows in the nuclei of radio galaxies are indeed well described by Bondi accretion in general, and (ii) that a significant fraction (about 1\%) of the rest-mass energy of the matter streaming into the Bondi radii of these systems ends up in jets.

Taking into account the optical AGN types, a remarkable pattern emerges: five out of six Seyfert 2 galaxies (the exception being UGC 6297) are located systematically below the best-fit linear relation between $\PB$ and $\Pj$, by about 0.8 dex. Noting that most of the remaining galaxies are actually located above the model line, one finds that the five Sy\,2s are separate from the remainder of the sample by about one order of magnitude in terms of jet production efficiency. Even though selection effects cannot be ruled out entirely given the limited sample size, this distribution indicates that Seyfert 2 galaxies are physically different from the other sources. This discrepancy is statistically significant: there are six out of 27 galaxies located below the model line by more than their measurement errors in both $\PB$ and $\Pj$, with five of them being Sy\,2s. When drawing randomly six out of 27 sources, the probability of obtaining a set with at least five Sy\,2s is $4.3\times10^{-4}$, corresponding to a Gaussian significance of $3.5\sigma$.

Within the frame of the standard viewing-angle unification scheme of AGN \citep{antonucci1993,urry1995}, the difference between type 1 (Sy\,1, LINER 1) and type 2 (Sy\,2, LINER 2) galaxies arises from geometry: in case of type 2 AGN, the view into the central engine is obscured by a surrounding extended dust torus, whereas type 1 sources are seen at inclinations that permit a direct view into the central accretion zone. In contrast to this simple geometric explanation, Figure~\ref{fig:powers} suggests that the optical type of a given AGN is partially controlled by its jet production efficiency. A possible \emph{evolutionary} description is the following: initially, Seyfert 2 galaxies are AGN which are relatively inefficient in powering jets and whose central regions are heavily obscured by dust. At some point, a sufficient -- roughly one order of magnitude -- increase in jet power, and thus in the release of energy in the central accretion zone, leads to two effects. Firstly, a partial dissolution of the surrounding dust torus. Depending on the viewing geometry, this may result in the transition of a Sy\,2 to a type 1 galaxy; indeed, indications for such an evolutionary sequence were already reported by \citet{tran2003}, \citet{koulouridis2006} and, more recently, by \citet{villarroel2014}. Secondly, shock-ionization of the gas within the galactic nucleus by the jet. As already noted by \citet{heckman1980}, this mechanism provides a natural explanation for the presence of the optical low-ionization emission lines characteristic for LINERs (see also \citealt{reynaldi2013,couto2013,riffel2014} for recent observational evidence for shock ionization by jets). Eventually, the scenario I propose implies an evolution from Seyfert 2s to LINERs and Seyfert 1 galaxies caused by an increase in jet production efficiency. I note that the rather clear separation of the Seyfert 2s from the rest of the sample suggests that the transition in efficiency happens relatively fast compared to the characteristic life times of galactic radio jets, meaning $\lesssim10^6$ years (cf., e.g., \citealt{tadhunter2012}).

When taking into account the apparent dependence of jet production efficiency on optical AGN type, the $\PB-\Pj$ correlation could actually be much tighter than usually assumed. Indeed, when removing the six Sy\,2 as well as the two outliers M 87 and NGC 4696 from the sample, I find an efficiency $\eta\approx1$\% and an intrinsic scatter about the best-fit model line \emph{consistent with zero}. A priori, some intrinsic scatter, as well as outliers, have to be expected because (i) the efficiency of accretion flows partially depends on the angular momentum of the infalling gas and the geometry and strength of the magnetic fields in the central accretion zone (e.g., \citealt{narayan2011}), (ii) AGN activity shows strong red-noise type temporal variability (e.g., \citealt{park2012,park2014,kim2013}), and (iii) a dependence of the AGN power on the physical properties of host galaxies, nearest-neighbor galaxies, and the large-scale background densities \citep{choi2009,hwang2012}. Those effects seem to be less influential in general than currently assumed once AGN evolution is taken into account properly.

%%% ACKNOWLEDGMENTS (IF ANY) %%%%%%%%%%%%%%%%%%%%%%%%%%%%%%%%%%%%%%%%

\acknowledgments

This work made use of the NASA/IPAC Extragalactic Database, the SIMBAD Astronomical Database, and the software package DPUSER developed and maintained by \name{Thomas Ott} at MPE Garching. I acknowledge financial support from the Korean National Research Foundation (NRF) via Basic Research Grant 2012-R1A1A2041387.

%%% APPENDICES (IF ANY) %%%%%%%%%%%%%%%%%%%%%%%%%%%%%%%%%%%%%%%%%%%%%

%\appendix
%\section{Appendix Title}

%%% CALL LIST OF REFERENCES (natbib STYLE) %%%%%%%%%%%%%%%%%%%%%%%%%%

\end{document}